\documentclass[journal=jacsat,manuscript=article]{achemso}
\usepackage{chemformula}
\usepackage[T1]{fontenc} 
\usepackage[version=4]{mhchem}
\usepackage{adjustbox}
\usepackage[colorlinks=true, allcolors=blue]{hyperref}

\author{Arijit Manna$^{1}$}
\email{amanna.astro@gmail.com}
\phone{+918777749445}
\affiliation{$^{1}$Department of Physics and Astronomy, Midnapore City College, Kuturia, Bhadutala, Paschim Medinipur, West Bengal 721129, India}
\author{Sabyasachi Pal$^{1}$}

\title[t-HC(S)SH towards NGC 1333 IRAS 4A2]{First Identification and Chemical Modeling of New Thiol (--SH) Bearing Molecule in the Interstellar Medium: Dithioformic Acid}
\keywords{astrochemistry, complex organic molecules, hot molecular cores, high-mass star formation region, ISM}

\begin{document}


\begin{abstract}
The study of complex organic molecules containing thiol (--SH) groups is essential in interstellar media because --SH plays an important role in the polymerization of amino acids (R-CH(NH$_{2}$)-COOH). Some quantum chemical studies have shown that there is a high chance of detecting the emission lines of dithioformic acid (HC(S)SH) in the highly dense and warm-inner regions of hot molecular cores and hot corinos. Therefore, we attempted to search for the emission lines of HC(S)SH toward the highly dense hot corino object NGC 1333 IRAS 4A using the Atacama Large Millimeter/Submillimeter Array (ALMA) band 7. We present the first detection of the rotational emission lines of the trans-conformer of dithioformic acid (t-HC(S)SH) toward the NGC 1333 IRAS 4A2. The column density and excitation temperature of the t-HC(S)SH toward NGC 1333 IRAS 4A2 are (2.63$\pm$0.32)$\times$10$^{15}$ cm$^{-2}$ and 255$\pm$32 K, respectively. The fractional abundance of t-HC(S)SH with respect to H$_{2}$ is (2.53$\pm$0.68)$\times$10$^{-9}$. The column density ratio of t-HC(S)SH and t-HCOOH toward NGC 1333 IRAS 4A2 is 0.36$\pm$0.02. To understand the possible formation pathways of HC(S)SH, we computed a two-phase warm-up chemical model abundance of HC(S)SH using the gas-grain chemical code UCLCHEM. After chemical modeling, we claim that HC(S)SH is formed in NGC 1333 IRAS 4A2 via barrierless radical--radical reactions between CSSH and H on the grain surfaces.
\end{abstract}
\text{\footnotesize keywords:}~\text{{\footnotesize astrochemistry, complex organic molecules, hot corinos, high-mass star formation region, ISM}}

\section{1. Introduction}
\label{sec:intro} 

More than three hundred molecules were detected at (sub)millimeter wavelengths, including the thiol (--SH)-bearing molecules methanethiol (\ce{CH3SH}), ethanethiol (\ce{C2H5SH}), and thioformic acid (HC(O)SH) in the interstellar medium (ISM) and circumstellar shells. Sulphur (S) is the 10$^{th}$ most abundant atomic compound in ISM. Among the other molecules, only twenty sulphur (S)-bearing molecules were detected in space. In ISM, the identified complex molecules with carbon (C), oxygen (O), hydrogen (H), and nitrogen (N) atoms carry a minimum of two atoms and a maximum of thirteen atoms, but most of S-bearing molecules have at most four atoms, such as thioformaldehyde (\ce{H2CS}) \citep{sin73}. This is because of the low cosmic abundance of atomic sulphur ($\sim$10$^{-5}$) with respect to \ce{H2}, which is less than ten times the abundance of O or C \citep{as09}. Sulphur chemistry has attracted interest since the early 1970s when the emission and absorption lines of carbon monosulfide (CS), carbonyl sulfide (OCS), methanethiol (\ce{CH3SH}), and hydrogen sulfide (\ce{H2S}) were detected in space \citep{pen71, je71, th72, lin79}. In diffuse clouds, sulphur is found in the form of S$^{+}$ in the gas phase, with a cosmic abundance of $\sim$10$^{-5}$ \citep{jen87, sav96}. Several previous studies have claimed that most of the missing sulphur is bound in the solid phase of dense regions in the form of refractory sulphur polymers (i.e., \ce{S8}) \citep{wak05, wak08} or polysulphanes (i.e., H$_{x}$S$_{y}$) \citep{dr12} but the primary sources of solid sulphur remains a mystery. Some chemical simulations show that carbon disulfide (\ce{CS2}) is a significant sink of sulphur on ice and dust analogues \citep{fer08, gar10}. Previously, the evidence of \ce{CS2} was found in some comets in the visible and UV regions \citep{jak04}. The emission lines of \ce{CS2} were also found in the coma of comet 67P/Churyumov-Gerasimenko \citep{cal16}, whose abundance was greater than that of the abundance of \ce{H2CS} towards Sgr B2 \citep{sin73}. Except for the comets, evidence of \ce{CS2} was not found in the interstellar ice. In shocked and warm regions, \ce{CS2} was transferred to the gas phase via grain evaporation \citep{pru18}. Therefore, the identification of \ce{CS2} in radio- and millimeter-wavelength telescopes is impossible because that molecule does not have any permanent dipole moments \citep{pru18}. Therefore, it is important to study other S-bearing molecules in the ISM.

\begin{figure}
	\centering
	\includegraphics[width=0.5\textwidth]{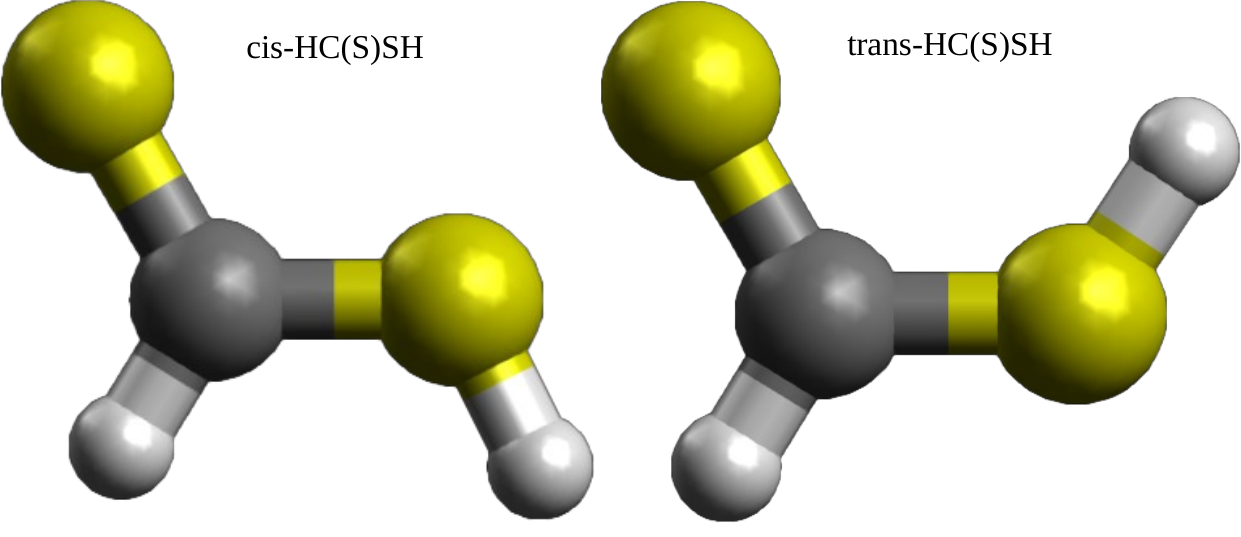}
	\caption{Three-dimensional molecular structure of cis- and trans-conformer of HC(S)SH.}
	\label{fig:molecular-structure}
\end{figure}

Many attempts have been made to improve our knowledge of the chemistry of interstellar sulphur by identifying new compounds, such as \ce{S3}, \ce{S4}, \ce{C2H5SH}, and t-HC(O)SH towards the coma of comet 67P/Churyumov-Gerasimenko \citep{cal16}, Orion KL \citep{kol14}, G+0.693--0.027 \citep{rod21}, G31.41+0.31 \citep{gar22}, and IRAS 16293--2422 B \citep{man23}. Moreover, the addition of these molecules, along with numerous other reactions and species, to the sulphur chemical networks has significantly enhanced modern chemical models \citep{wo15, vi17}. Despite these developments, several significant compounds, such as dithioformic acid (HC(S)SH), are still overlooked in the current understanding of sulphur chemistry. This molecule is the S-bearing counterpart of formic acid (HCOOH), and it is chemically linked to \ce{CS2} \citep{pru18}. Since HC(S)SH and HCOOH are isostructural, they can exist in two distinct conformations (cis and trans), depending on how the S-H bond is oriented in relation to the C-H bond \citep{pru18}. The molecular structures of the cis- and trans-conformers of HC(S)SH are shown in Figure~\ref{fig:molecular-structure}. The cis- and trans-conformers of HC(S)SH are near-prolate asymmetric rotors with planar structures and $C_{s}$ symmetry ($k_{cis} = -0.968$ and $k_{trans} = -0.990$) \citep{pru18}. The dipole moments of a component of cis- and trans-conformers of HC(S)SH are $\mu_{a}$ (cis) = 2.10 Debye and $\mu_{a}$ (trans) = 1.53 Debye \citep{bak79}. Previous quantum chemical modelling has shown that the trans-conformer (hereafter t-HC(S)SH) is the most stable, with an energy difference of 421 cm$^{-1}$ with respect to the cis-conformer (hereafter c-HC(S)SH) of HC(S)SH \citep{pru18}. The spectroscopic parameters of t-HC(S)SH and c-HC(S)SH are described in detail in \citet{pru18} and \citet{bak79}. Previously, scientists attempted to detect the emission lines of t-HC(S)SH towards Orion KL in the frequency ranges of 86 GHz and 279 GHz using IRAM without any success owing to the lower angular and spectral resolution of this telescope \citep{es13}. The upper-limit column density of t-HC(S)SH towards Orion KL is $\leq$3.6$\times$10$^{13}$ cm$^{-2}$ \citep{es13}. Quantum chemical studies show that there is a high chance of detecting the emission lines of t-HC(S)SH towards warm regions in ISM above 300 GHz \citep{pru18}. The majority of sulphur-bearing molecules, such as CS, NS, NS$^{+}$, SO, SO$^{+}$, \ce{H2S}, HDS, \ce{D2S}, HCS$^{+}$, CCS, OCS, \ce{SO2}, \ce{H2CS}, HDCS, \ce{D2CS}, HSCN, CCCS, and \ce{CH3SH}, were detected towards NGC 1333 IRAS 4A2 \citep{taq20, qui24}, and previous chemical modelling of \citet{kou17} showed that NGC 1333 IRAS 4A2 is an ideal source for studying new S-bearing molecules. Therefore, we attempted to search for the absorption and emission lines of t-HC(S)SH towards NGC 1333 IRAS 4A1 (hereafter A1) and NGC 1333 IRAS 4A2 (hereafter A2) in the frequency ranges of 349.75 GHz and 364.33 GHz. We study the absorption and emission lines of t-HC(S)SH towards A1 and A2 in the frequency ranges of 349.75 GHz and 364.33 GHz because the quantum chemical modelling of \citet{pru18} showed that most of all high-intensity transitions of t-HC(S)SH will be detected above 350 GHz with higher upper state energies ($>$ 400 K). Earlier, \citet{pru18} also showed that 54$_{7*}$--53$_{7*}$ is an efficient transition of t-HC(S)SH and that these types of transitions are found in the frequency ranges of 349.75 GHz and 364.33 GHz. The description of the A1 and A2 is already discussed in \citet{sah19}.

\begin{figure*}[!ht]
	\centering
	\centering
	\includegraphics[width=1.0\textwidth]{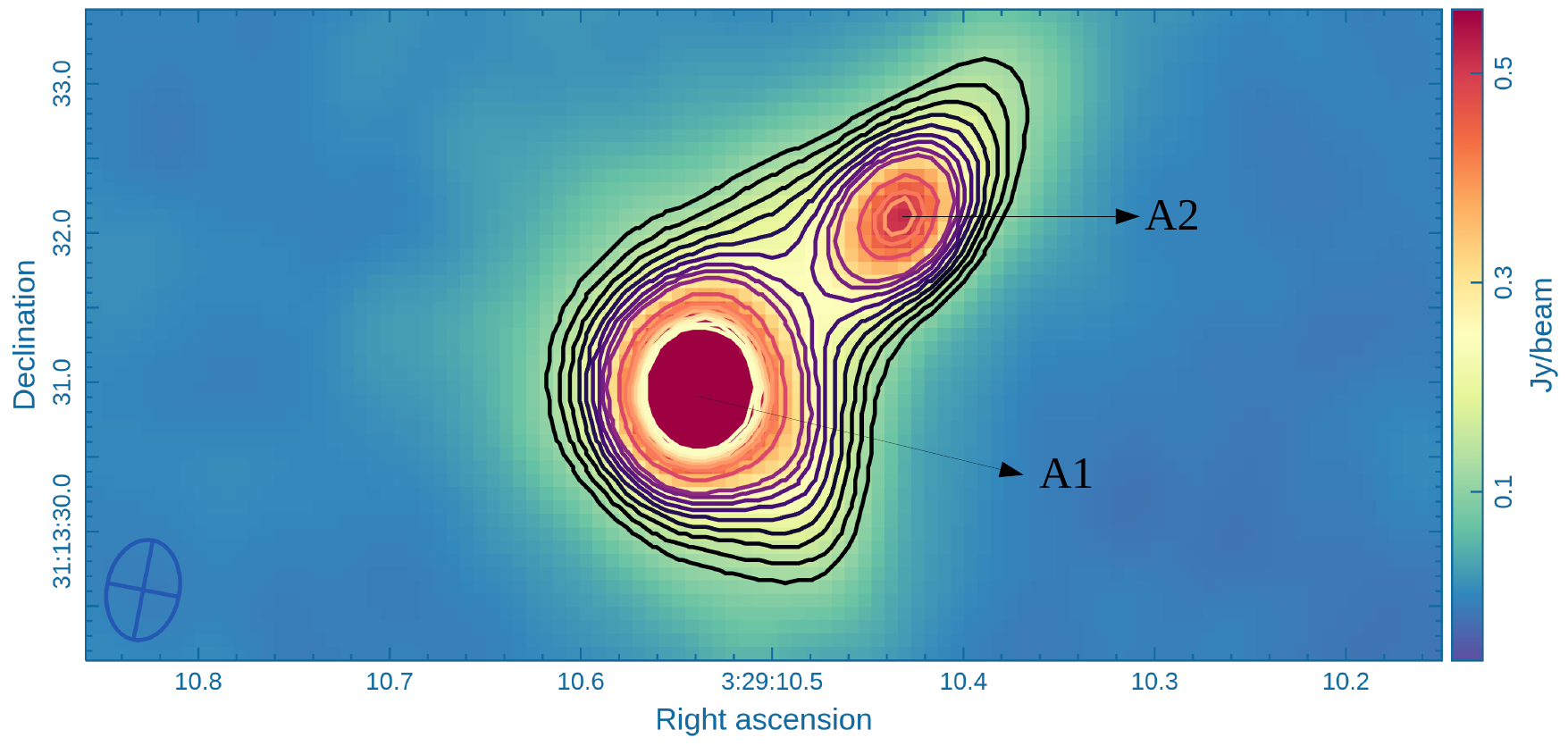}
	\caption{The continuum emission image of NGC 1333 IRAS 4A at a frequency of 350.86 GHz (0.85 mm). The contour levels start at 3$\sigma$, where $\sigma$ is the root mean square (RMS) of the continuum image. The contour levels increased by a factor of $\surd$2. The blue circle indicates the synthesized beam of the continuum image.}
	\label{fig:continuum}
\end{figure*}

\begin{figure*}[!ht]
	\centering
	\centering
	\includegraphics[width=1.0\textwidth]{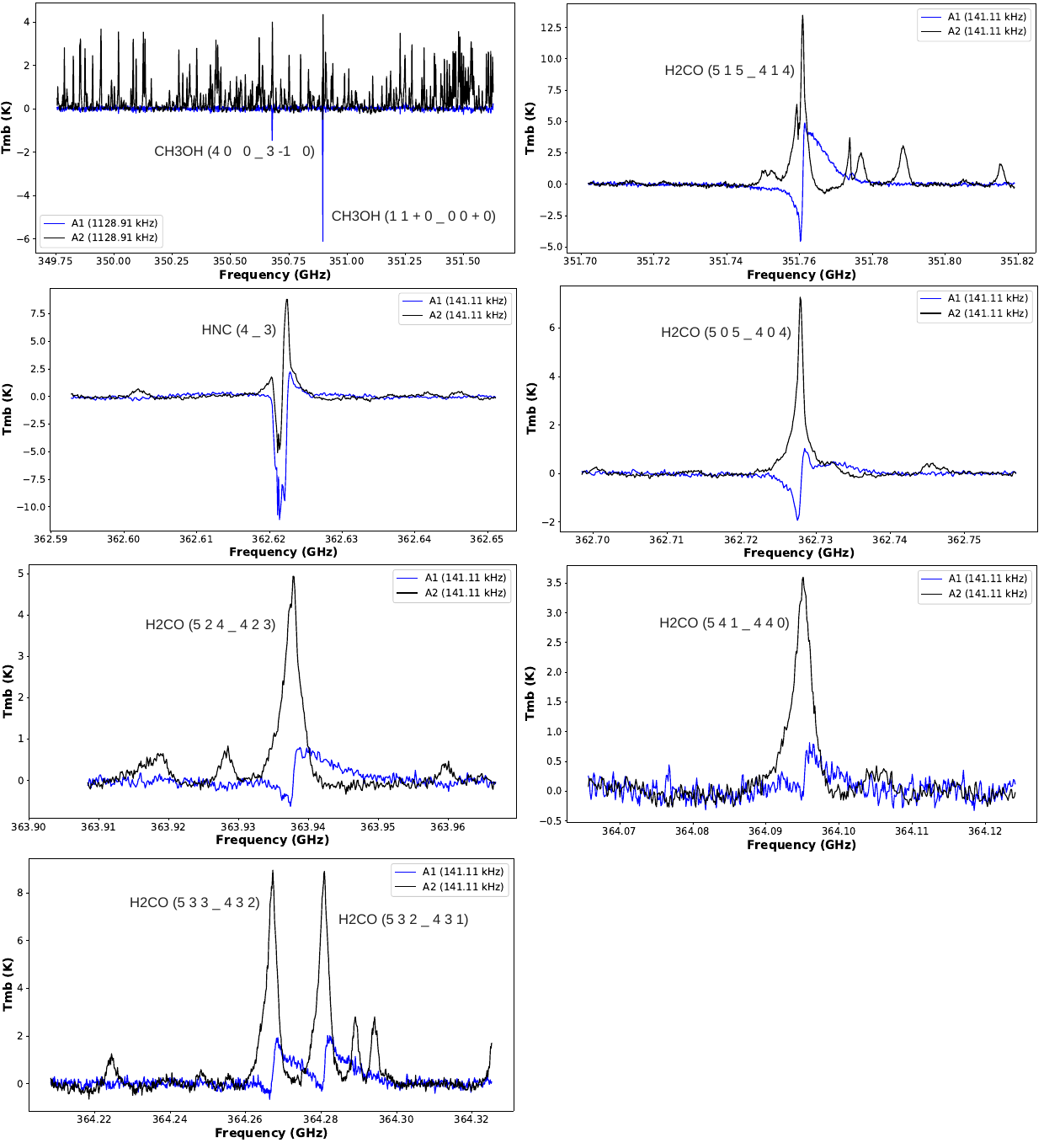}
	\caption{Molecular spectra of A1 and A2 between the frequency ranges of 349.75--351.63 GHz, 351.70--351.82 GHz, 362.59--362.65 GHz, 362.70--362.76 GHz, 363.91--363.97 GHz, 364.07--364.12 GHz, and 364.21--364.33 GHz. The resolution of each spectrum is shown in each panel of the spectra. The blue lines indicate the molecular absorption spectra of A1 and the black lines represent the molecular emission spectra of A2.}
	\label{fig:fullspectra}
\end{figure*}

\begin{figure*}
	\centering
	\includegraphics[width=1.0\textwidth]{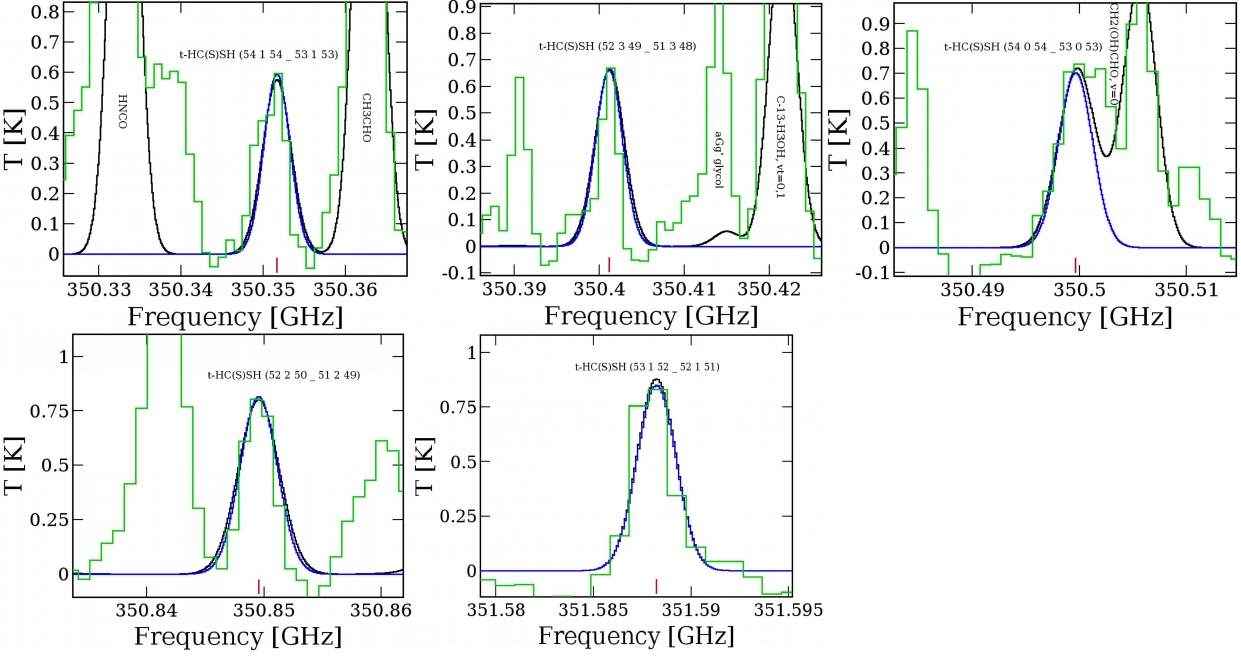}
	\caption{Rotational emission lines of t-HC(S)SH towards A2 in the frequency ranges of 349.75 GHz and 351.63 GHz. The green lines show the molecular spectra of A2, the blue lines show the LTE model spectra of t-HC(S)SH, and the black lines indicate the LTE spectra of the other molecules, including t-HC(S)SH.}
	\label{fig:line}
\end{figure*}

\section{2. Previous observation and data analysis}
We used the archival data of the NGC 1333 IRAS 4A, which was observed using the high-resolution Atacama Large Millimeter/submillimeter Array (ALMA) band 7 with 12-m arrays (PI: Su, Yu-Nung, 2015.1.00147.S). The phase centre of NGC 1333 IRAS 4A is ($\alpha,\delta$)$_{\rm J2000}$ = 03:29:10.500, +31:13:31.500. This observation was performed on July 23, 2016, using 39 antennas with an on-source integration time of 3326.40 s. At the time of observation, the minimum and maximum baseline were 15.4 m and 1100 m, respectively. During the observation, the flux and bandpass calibrators were taken as J0510+1800 and J0006--0623, respectively. The phase calibrator was set as J0336+3218. This observation was made in the frequency ranges of 349.75--364.33 GHz with a spectral resolution of 1128.91 kHz and 141.11 kHz.

For data reduction and imaging, we used the Common Astronomy Software Application (CASA 5.4.1) with an ALMA data-reduction automated pipeline \citep{mc07}. Initially, we used the Perley-Butler 2017 flux calibrator model for each baseline to scale the continuum flux density of the flux calibrator using the CASA task SETJY \citep{per17}. We created flux and bandpass calibration after flagging the defective channels and antenna data using the CASA pipeline tasks HIFA\_BANDPASSFLAG and HIFA\_FLAGDATA. Subsequently, we used the task MSTRANSFORM with all available rest frequencies to split the target data of the NGC 1333 IRAS 4A. We produced continuum emission maps of the NGC 1333 IRAS 4A for line-free channels using the CASA task TCLEAN with the HOGBOM deconvolver. We used the UVCONTSUB task in the UV plane of the separated calibrated data for the continuum subtraction. Subsequently, we created spectral data cubes of NGC 1333 IRAS 4A using the CASA task TCLEAN with the SPECMODE = CUBE parameter. Finally, we used the CASA task IMPBCOR to correct the synthesized beam pattern in the continuum and spectral data cubes of NGC 1333 IRAS 4A.

\begin{table*}
	\centering 
	\caption{Summary of the molecular line properties of t-HC(S)SH towards A2.}
	\begin{adjustbox}{width=1.0\textwidth}
		\begin{tabular}{ccccccccccccccccccc}
			\hline 
			Frequency &Transition&$E_{u}$ & $A_{ij}$ &g$_{up}$&FWHM&$\rm{\int T_{mb}dV}$&Optical depth&Remarks\\
			(GHz)&(${\rm J^{'}_{K_a^{'}K_c^{'}}}$--${\rm J^{''}_{K_a^{''}K_c^{''}}}$)&(K)&(s$^{-1}$)& &(km s$^{-1}$)&(K km s$^{-1}$)  &($\tau$)& \\
			\hline
			350.351&54(1,54)--53(1,53)&466.43&5.40$\times$10$^{-4}$&109&3.20$\pm$0.82&1.86$\pm$0.20 &3.69$\times$10$^{-3}$&Non blended\\
			
			350.401&52(3,49)--51(3,48)&463.02&5.39$\times$10$^{-4}$&105&3.20$\pm$0.29&1.72$\pm$0.21&3.59$\times$10$^{-3}$&Non blended\\
			
			350.499&54(0,54)--53(0,53)&466.37&5.41$\times$10$^{-4}$ &109&3.20$\pm$0.35&--&3.52$\times$10$^{-3}$&Blended with \ce{CH2(OH)CHO}\\
			
			350.849&52(2,50)--51(2,49)&454.60&5.24$\times$10$^{-4}$&105&3.20$\pm$0.25&1.92$\pm$0.29&3.72$\times$10$^{-3}$&Non blended\\
			
			351.588&53(1,52)--52(1,51)&462.87&5.46$\times$10$^{-4}$&107&3.20$\pm$0.89&--&3.68$\times$10$^{-3}$&Blended with \ce{CH3CHO}\\
			
			\hline
		\end{tabular}	
	\end{adjustbox}
	\label{tab:MOLECULAR DATA}\\
\end{table*}

\section{3. Results}
\subsection{3.1 Continuum emission towards NGC 1333 IRAS 4A}
The dust continuum emission image of NGC 1333 IRAS 4A at a frequency of 350.86 GHz (0.85 mm) is shown in Figure~\ref{fig:continuum}. NGC 1333 IRAS 4A contains two hot corinos (A1 and A2) separated by a distance of 1.8$^{\prime\prime}$ from each other. The synthesized beam size of the continuum emission image was 0.68$^{\prime\prime}$$\times$0.49$^{\prime\prime}$. To determine the physical properties of the dust continuum emission image, we fitted a 2D Gaussian using the task IMFIT over A1 and A2. The integrated and peak flux densities of A1 were 1.47$\pm$0.04 Jy and 744$\pm$15 mJy beam$^{-1}$, with an RMS of 21 mJy. The integrated and peak flux densities of A2 were 723$\pm$59 mJy and 398$\pm$22 mJy beam$^{-1}$, with an RMS of 20 mJy. The deconvolved source sizes of A1 and A2 were 0.82$^{\prime\prime}$$\times$0.52$^{\prime\prime}$ and 0.70$^{\prime\prime}$$\times$0.31$^{\prime\prime}$. The deconvolved source sizes of A1 and A2 were larger than the synthesized beam size of the continuum image, which means that the continuum emission map of NGC 1333 IRAS 4A was resolved.

\begin{figure}
	\centering
	\includegraphics[width=1.0\textwidth]{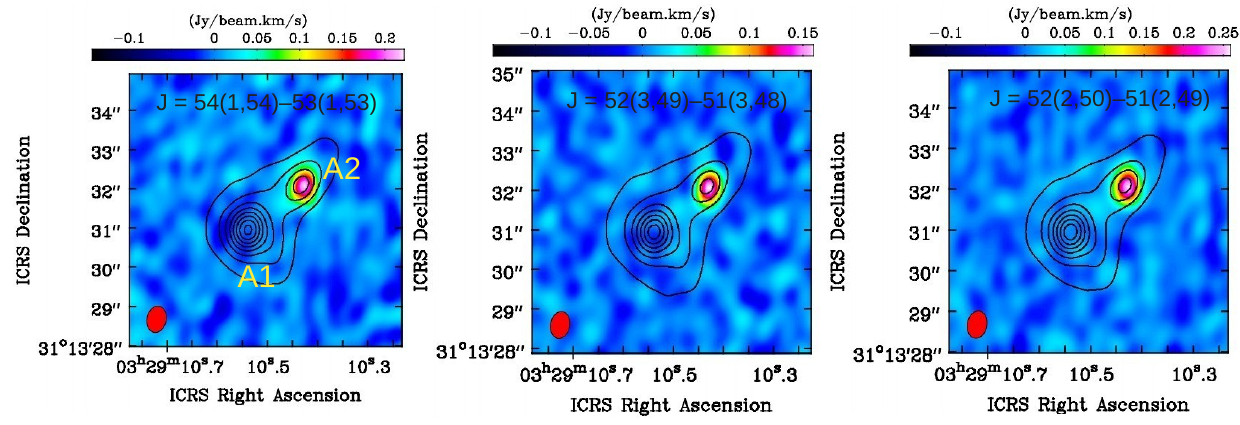}
	\caption{Integrated emission maps of non-blended transitions of t-HC(S)SH towards A2, which are overlaid with the 0.85 mm continuum emission map (black contours). The contour levels began at 3$\sigma$. The red circle indicates the synthesized beam of the integrated emission maps.}
	\label{fig:emissionmap}
\end{figure}

\subsubsection{3.1.1 Estimation of column density of molecular \ce{H2}}
The peak flux density ($S_\nu$) for optically thin dust continuum emission can be described as  
	
\begin{equation} 
S_\nu = B_\nu(T_d)\tau_\nu\Omega_{beam} 
\end{equation} 
In the above equation, $\tau_\nu$ indicates the optical depth, $B_\nu(T_d)$ indicates the Planck function at dust temperature $T_d$ \citep{whi92}, and $\Omega_{beam} = (\pi/4 \ln 2)\times \theta_{major} \times \theta_{minor}$ indicates the solid angle of the synthesized beam. The equation of optical depth in terms of the mass density of dust can be expressed as,  
\begin{equation} 	
\tau_\nu =\rho_d\kappa_\nu L 
\end{equation} 
where $\kappa_{\nu}$ is the mass absorption coefficient, $L$ is the path length, and $\rho_d$ is the mass density of dust. The mass density of the dust can be shown in terms of the dust-to-gas mass ratio ($Z$),  
\begin{equation} 	
\rho_d = Z\mu_H\rho_{H_2}=Z\mu_HN_{H_2}2m_H/L 
\end{equation} 
where $\rho_{H_2}$ is the hydrogen mass density, $\mu_H$ is the mean atomic mass per hydrogen atom, $m_H$ is the mass of hydrogen, and the column density of molecular \ce{H2} can be expressed as $N_{H_2}$. We use $\mu_H = 1.41$, $Z = 0.01$ \citep{cox00}, and $T_d$ = 60 K \citep{sah19, taq15}. The peak flux densities of A1 and A2 were 744$\pm$15 mJy beam$^{-1}$ and 398$\pm$22 mJy beam$^{-1}$, respectively. According to the equations 1, 2, and 3, the equation of the column density of molecular \ce{H2} can be expressed as,  
\begin{equation} 	
N_{H_2} = \frac{S_\nu /\Omega}{2\kappa_\nu B_\nu(T_d)Z\mu_H m_H} 
\end{equation} 
To estimate the value of $\kappa_{\nu}$, we adopted the formula $\kappa_\nu = 0.90(\nu/230~\textrm{GHz})^{\beta}\ \textrm{cm}^{2}\ \textrm{g}^{-1}$ \citep{moto19}, where $k_{230} = 0.90$ cm$^{2}$ g$^{-1}$ indicates the emissivity of dust grains at a gas density of $\rm{10^{6}\ cm^{-3}}$ covered by a thin ice mantle at 230 GHz. We use the dust spectral index $\beta$ $\sim$ 1.5 \citep{taq15}. The estimated value of $\kappa_{\nu}$ was 1.695. The estimated column densities of molecular \ce{H2} towards A1 and A2 using equation 4 were (1.89$\pm$0.45)$\times$10$^{24}$ cm$^{-2}$ and (1.04$\pm$0.25)$\times$10$^{24}$ cm$^{-2}$.

\subsection{3.2 Line emission towards NGC 1333 IRAS 4A}
First, we extracted molecular spectra from the spectral data cubes of NGC 1333 IRAS 4A to draw 1.5$^{\prime\prime}$ diameter circle over the A1 and A2 regions. The synthesized beam sizes of the spectral data cubes between the frequency ranges of 349.75 GHz and 364.33 GHz vary between 0.68$^{\prime\prime}$$\times$0.49$^{\prime\prime}$ and 0.69$^{\prime\prime}$$\times$0.50$^{\prime\prime}$. We observed that the first spectral data cube whose frequency range between 349.75 GHz and 351.63 GHz showed rich molecular spectra. The other spectral windows data cubes between the frequency ranges of 351.70--351.82 GHz, 362.59--362.65 GHz, 362.70--362.76 GHz, 363.91--363.97 GHz, 364.07--364.12 GHz, and 364.21--364.33 GHz show the molecular outflows such as \ce{H2CO} and HNC, which were already reported by other authors \citep{su19}. The observed spectra of A1 and A2 between the frequency ranges of 349.75 GHz and 364.33 GHz are shown in Figure~\ref{fig:fullspectra}. The extracted molecular spectra of A2 showed molecular emission lines, whereas the molecular spectra of A1 showed absorption features. This occurred because the dust emission in A1 was optically thick, whereas that in A2 was optically thin. The systematic velocities ($V_{LSR}$) of A1 and A2 are 7.03 km s$^{-1}$ and 6.30 km s$^{-1}$, respectively \citep{sah19,su19}.

\subsubsection{3.2.1 Detection of t-HC(S)SH towards NGC 1333 IRAS 4A}
We focused on studying the absorption and emission lines of t-HC(S)SH towards A1 and A2 because t-HC(S)SH is the most stable, with an energy difference of 421 cm$^{-1}$ with respect to c-HC(S)SH, and there is a high chance of detecting that molecule in the warm-inner region of A1 and A2. We study the absorption and emission lines of t-HC(S)SH from the spectra of A1 and A2 in the frequency ranges of 349.75--351.63 GHz because those spectra have a 1.88 GHz spectral span (see first spectra in Figure~\ref{fig:fullspectra}). The other spectra between the frequency ranges of 351.70--351.82 GHz, 362.59--362.65 GHz, 362.70--362.76 GHz, 363.91--363.97 GHz, 364.07--364.12 GHz, and 364.21--364.33 GHz have 0.12 GHz, 0.06 GHz, 0.06 GHz, 0.06 GHz, 0.05 GHz, and 0.12 GHz spectral spans. Due to insufficient spectral spans, we only identified the outflows of different transitions of \ce{H2CO} and HNC rather than any complex molecules between the frequency ranges of 351.70--351.82 GHz, 362.59--362.65 GHz, 362.70--362.76 GHz, 363.91--363.97 GHz, 364.07--364.12 GHz, and 364.21--364.33 GHz, respectively. Earlier, \citet{sah19} and \citet{sahu20} also used this data and found the emission lines of \ce{CH3OH}, $^{13}$\ce{CH3OH}, \ce{CH2DOH}, \ce{CH3CHO}, and \ce{CH3OC(O)NH2} from the spectra of A2 between the frequency range of 349.75--351.63 GHz. Previous studies showed that \ce{CH3OC(O)NH2} is an isomer of the simplest amino acid glycine (\ce{NH2CH2COOH}) \citep{sahu20}. 

To detect the absorption and emission lines of t-HC(S)SH towards A1 and A2, we used the local thermodynamic equilibrium (LTE) model with the Cologne Database for Molecular Spectroscopy (CDMS entry 078506) molecular database \citep{mu05}. The LTE-RADEX module in CASSIS is used to fit the LTE spectra to the observed spectra of t-HC(S)SH \citep{vas15}. The gas densities of the warm inner regions of A1 and A2 are 2.5$\times$10$^{8}$ cm$^{-3}$ and 1.3$\times$10$^{8}$ cm$^{-3}$, suggesting that the LTE assumptions are appropriate \citep{taq15}. During the LTE spectral modelling of t-HC(S)SH, we considered a background continuum source temperature of $T_{c}$ = 0 and a cosmic background temperature of $T_{bg}$ = 2.73 K. After LTE spectral modelling, we identified five high-intensity transition lines of t-HC(S)SH between the frequency ranges of 349.75 GHz and 351.63 GHz towards A2. The upper-level energies ($E_{u}$) of the five detected transition lines of t-HC(S)SH vary between 454.60 K and 466.43 K, where the Einstain coefficient ($A_{ij}$) varies between 5.24$\times$10$^{-4}$ s$^{-1}$ and 5.46$\times$10$^{-4}$ s$^{-1}$. To understand the blended effect, we also fitted more than 250 molecular transitions, including the rotational emission lines of \ce{CH3OH}, $^{13}$\ce{CH3OH}, \ce{CH2DOH}, and \ce{CH3CHO}, which have already been reported by other authors using this data \citep{sah19}. The LTE-fitted rotational emission lines of the t-HC(S)SH towards A2 are shown in Figure~\ref{fig:line}. After spectral analysis, we see that the $J$ = 54(0,54)--53(0,53) and $J$ = 53(1,52)--52(1,51) transition lines of t-HC(S)SH are blended with the simplest sugar-like molecule \ce{CH2OHCHO} and \ce{CH3CHO}. The other three non-blended lines exhibited above 4$\sigma$ significance. There are no missing transition lines of t-HC(S)SH between the frequency ranges of 349.75 GHz and 351.63 GHz. The spectral line parameters of the t-HC(S)SH for A2 are listed in Table~\ref{tab:MOLECULAR DATA}. Based on LTE spectral modelling, the best-fit column density of t-HC(S)SH towards A2 is (2.63$\pm$0.32)$\times$10$^{15}$ cm$^{-2}$ with an excitation temperature of 255$\pm$32 K and a source size of 0.68$^{\prime\prime}$. The estimated excitation temperature of t-HC(S)SH was similar to that of \ce{CH3CHO} towards A2, which was estimated by \citet{sah19}. The full-width half maximum (FWHM) of the LTE model spectra of t-HC(S)SH is 3.20 km s$^{-1}$. The fractional abundance of t-HC(S)SH with respect to H$_{2}$ towards A2 was (2.53$\pm$0.68)$\times$10$^{-9}$, where the column density of molecular H$_{2}$ towards A2 was (1.04$\pm$0.25)$\times$10$^{24}$ cm$^{-2}$. Previous studies have shown that the column density and abundance of t-HCOOH towards A2 were 7.3$\times$10$^{15}$ cm$^{-2}$ and 4.6$\times$10$^{-9}$ \citep{bot04}, which is similar to the abundance of t-HC(S)SH towards A2. Since HC(S)SH and HC(O)SH are counterparts of HCOOH, the column density ratio of t-HC(S)SH and t-HCOOH towards A2 is 0.36$\pm$0.02, which is very close to that of the column density ratio of t-HC(O)SH and t-HCOOH towards the hot molecular core G31.41+0.31 \citep{gar22}. After detection of the emission lines of t-HC(S)SH towards A2, we also searched the absorption lines of t-HC(S)SH towards A1 using the LTE model spectra, but we could not identify them. The upper-limit column density and fractional abundance of t-HC(S)SH towards A1 are $\leq$(3.58$\pm$0.59)$\times$10$^{13}$ cm$^{-2}$ and $\leq$(1.89$\pm$0.92)$\times$10$^{-11}$, respectively.

\subsection{3.3 Spatial distribution t-HC(S)SH}
After identifying the rotational emission lines of t-HC(S)SH towards A2, we created integrated emission maps of three non-blended transitions of t-HC(S)SH by integrating the spectral data cubes in the velocity ranges (5--9 km$^{-1}$, 4--8 km$^{-1}$, and 4--8 km$^{-1}$), where the emission lines of t-HC(S)SH were detected. The integrated emission maps of t-HC(S)SH were overlaid with a 0.85 mm continuum emission image of the NGC 1333 IRAS 4A. The integrated emission maps of t-HC(S)SH are presented in Figure~\ref{fig:emissionmap}. The synthesized beam sizes of the integrated emission maps are 0.68$^{\prime\prime}\times$0.49$^{\prime\prime}$, 0.69$^{\prime\prime}\times$0.47$^{\prime\prime}$, and 0.68$^{\prime\prime}\times$0.48$^{\prime\prime}$. Based on the integrated emission maps, we found that the t-HC(S)SH originated from the warm-inner regions of A2. By fitting 2D Gaussians over the integrated emission maps, we estimated the sizes of the emitting regions of t-HC(S)SH. To determine the deconvolved beam size of the emitting region of t-HC(S)SH, we used the following equation \citep{man23a, man24}:\\

\begin{equation}
	\theta_{S}=\sqrt{\theta^2_{50}-\theta^2_{beam}}
\end{equation}
In the above equation, $\theta_{50} = 2\sqrt{A/\pi}$ is the diameter of the circle whose area ($A$) is enclosed by the 50\% line peak and $\theta_{beam}$ is the half-power width of the synthesized beam of the integrated emission maps of t-HC(S)SH. The sizes of the emitting regions of t-HC(S)SH corresponding to the transition lines of $J$ = 54(1,54)--53(1,53), $J$ = 52(3,49)--51(3,48), and $J$ = 52(2,50)--51(2,49) were 0.66$^{\prime\prime}$, 0.67$^{\prime\prime}$, and 0.66$^{\prime\prime}$, respectively. The sizes of the emitting regions of t-HC(S)SH were smaller than the synthesized beam sizes of the integrated emission maps of t-HC(S)SH. This indicates that the integrated emission maps of t-HC(S)SH were not spatially resolved towards A2. Therefore, the chemical morphology of t-HC(S)SH cannot be determined from the integrated emission maps.

\begin{figure*}[!ht]
	\centering
	\centering
	\includegraphics[width=1.0\textwidth]{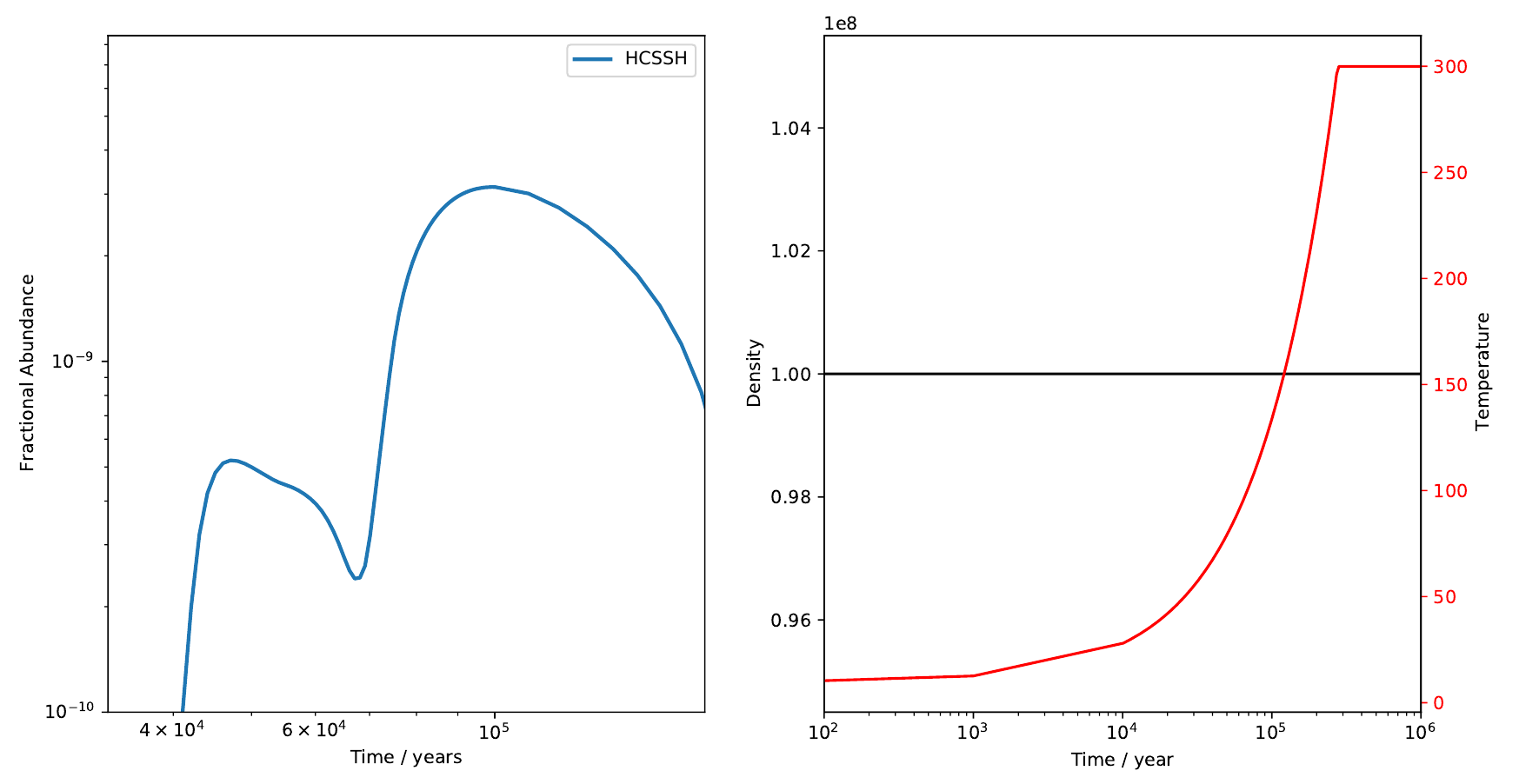}
	\caption{Two-phase warm-up chemical model abundance of HC(S)SH with respect to time on the grain surface of hot corinos (left panel). The red line indicates the temperature profile of the chemical model (right panel).}
	\label{fig:model}
\end{figure*}

\section{4. Discussions}
\subsection{4.1 Chemical modelling of HC(S)SH}
To determine the modelled abundance and possible formation mechanism of HC(S)SH, we computed a two-phase (gas + grain) warm-up chemical model of HC(S)SH using the time-dependent gas-grain chemical code UCLCHEM \citep{hol17}. The UCLCHEM chemical code focuses on grain surface chemistry as well as gas phase reactions towards hot corinos, hot molecular cores, and molecular clouds \citep{hol17}. Thermal and non-thermal desorption, grain surface, and gas-phase reaction networks are included in this code. UCLCHEM solves the reaction rates and determines the fractional abundances of molecules with respect to time on the grain surface and gas phase \citep{hol17}. In the two-phase warm-up chemical model, the free-fall collapse of the cloud (Phase I) was followed by a warm-up phase (Phase II). In the first phase (Phase I), the gas density increased from $n_{H}$ = 1$\times$10$^{2}$ cm$^{-3}$ to 1$\times$10$^{8}$ cm$^{-3}$, and the gas and grain temperatures remained constant at 10 K. During chemical modelling, the initial visual extinction ($A_{V}$) was 2 and the cosmic ray ionization rate was 1.3$\times$10$^{-17}$ s$^{-1}$. During this time, atoms and molecules are accreted on the grain surface at an accretion rate of 10$^{-5}$ \textup{M}$_{\odot}$ yr$^{-1}$, and this accretion rate depends on the gas density of the hot corinos \citep{vi04}. During chemical modelling, we assumed the sticking probability to be unity, which means that all incoming hydrogen atoms stick to the grain surface if they get an empty site. During this time, the molecular species may hydrogenate or rapidly react with other species on the grain surface. During the chemical modelling of HC(S)SH, the abundances of atomic elements, such as oxygen (O), carbon (C), nitrogen (N), and helium (He), corresponding to the solar values, were obtained from another article \citep{as09}. In chemical models, other atomic elements, such as silicon (Si), sulphur (S), chlorine (Cl), magnesium (Mg), phosphorus (P), and fluorine (F) are depleted by factors of 100. In the second phase (phase II), the gas density was fixed at 1$\times$10$^{8}$ cm$^{-3}$, but the temperature was increased from 10 to 300 K. Phase II is known as the warm-up stage of hot corinos, in which the temperature increases according to the formula $T = T_0 + (T_{max}-T_0)(\Delta t/t_h)^n$ \citep{gar13}. In the warm-up stage, the gas and dust temperatures were assumed to be well-coupled. In the warm-up stage, the molecules no longer freeze, and the frozen molecules on the grain surface are left in the gas phase by both thermal and non-thermal desorption mechanisms. In addition, we included co-desorption with \ce{H2O}, volcano desorption, monomolecular desorption, and grain mantle desorption in our model \citep{col04, cou18}. This chemical modelling of HC(S)SH is similar to the two-phase warm-up chemical modelling of \citet{man24a} and \citet{man23b}.
	
We have added the following reactions of HC(S)SH in the chemical network of UCLCHEM: \\ 
HS + CS $\rightarrow$ CSSH~~~~~~~~~~~~~~~~~~~~~~~~~~~~~(Reaction 1)\\ 
CSSH + H $\rightarrow$ HC(S)SH ~~~~~~~~~~~~~~~~~~~~~(Reaction 2)\\ 
HS + \ce{H2CS} $\rightarrow$ HC(S)SH ~~~~~~~~~~~~~~~~~~~~(Reaction 3)\\ 
\ce{HCSSH2}$^{+}$ + e$^{-}$ $\rightarrow$ HC(S)SH + H ~~~~~~~~~(Reaction 4)\\ 
HC(S)SH $\rightarrow$ \ce{H2} + \ce{CS2} ~~~~~~~~~~~~~~~~~~~~~~~(Reaction 5)\\ 
Reactions 1 and 2 show that the reactions between HS and CS form CSSH on the grain surface, and barrierless radical-radical reactions between CSSH and H produce HC(S)SH on the grain surface \citep{pru18, las19}. Previous quantum chemical studies and a three-stage evolutionary cloud model have shown that reaction 2 is the most efficient for the formation of HC(S)SH towards the warmer part of the star-formation regions and molecular clouds \citep{pru18, las19}. Reaction 3 shows that the reaction between the HS and \ce{H2CS} produces HC(S)SH in the gas phase \citep{las19}. Reactions 2 and 3 are the most efficient reactions for the formation of HC(S)SH because the precursors of both reactions are CS and \ce{H2CS}, and both molecules were earlier detected towards A2 \citep{qui24}. Reaction 4 shows that the dissociative recombination of \ce{HCSSH2}$^{+}$ produces HC(S)SH in the gas phase \citep{las19}. Reaction 5 shows that HC(S)SH is destroyed via unimolecular decomposition and forms \ce{H2} and \ce{CS2} \citep{xi96}. We did not find any other chemical reactions of HC(S)SH in the KIDA and UMIST 2012 astrochemistry reaction network databases. After chemical modelling, we observed that reaction 2 produces highly abundant HC(S)SH on the grain surface. We also observed that reactions 3 and 4 do not produce HC(S)SH in the gas phase. Similarly, we also observed that reaction 5 has a 90\% ability to destroy HC(S)SH. The computed two-phase warm-up chemical model of HC(S)SH is shown in Figure~\ref{fig:model}. The maximum modelled abundance of HC(S)SH reached 3.17$\times$10$^{-9}$ on the grain surface at the time of 1$\times$10$^{5}$ yr, which is obtained from reaction 2 in the warm-up phase. 

To understand the possible formation pathways of HC(S)SH towards A2, we compared our estimated abundance of t-HC(S)SH with the modelled abundance of HC(S)SH, which was estimated by two-phase warm-up chemical modelling using the UCLCHEM. This comparison is physically reasonable because the gas density of A2 is 1.3$\times$10$^{8}$ cm$^{-3}$ \citep{taq15}. Hence, our computed two-phase warm-up chemical model based on timescales is appropriate for explaining the chemical evolution of HC(S)SH towards A2. From this observation, we found that the fractional abundance of t-HC(S)SH towards A2 is (2.53$\pm$0.68)$\times$10$^{-8}$, which is similar to the two-phase warm-up chemical model abundance of HC(S)SH. This comparison indicates that t-HC(S)SH is formed in A2 via barrierless radical-radical reactions between CSSH and H on grain surfaces.

\begin{table*}
	\centering
	\caption{Comparison of t-HC(S)SH abundance with that of other S-bearing molecules towards A2.}
	\begin{adjustbox}{width=0.85\textwidth}
		\begin{tabular}{|c|c|c|c|c|c|c|c|}
			\hline 
			Molecule&Column density ($N$)&Excitation temperature ($T_{ex}$)& Fractional abundance&Comments &References\\
			&(cm$^{-2}$)   &  (K)                 &  $X$ = (N$_{molecule}$/N$_{H_{2}}$)                   &        &   \\
			\hline
SiO&0.41$\times$10$^{12}$&7.9$\pm$1.6&4.1$\times$10$^{-11}$ &Cold component  &\citet{qui24}\\	
SO &88$\times$10$^{12}$&10.8$\pm$0.6&8.8$\times$10$^{-9}$ &Cold component  &\citet{qui24}\\	
SO$^{+}$&5.4$\times$10$^{12}$&9.8$\pm$0.6&5.4$\times$10$^{-10}$&Cold component  &\citet{qui24}\\
$^{33}$SO&2.6$\times$10$^{12}$&20.2$\pm$1.5&2.6$\times$10$^{-10}$&Cold component  &\citet{qui24}\\
S$^{18}$O&1.6$\times$10$^{12}$&8.7$\pm$2.7&1.6$\times$10$^{-10}$&Cold component  &\citet{qui24}\\
$^{34}$SO&8.3$\times$10$^{12}$&10.7$\pm$0.8&8.3$\times$10$^{-10}$&Cold component  &\citet{qui24}\\
OCS&26$\times$10$^{12}$&72.8$\pm$18.3&2.6$\times$10$^{-9}$&Warm component  &\citet{qui24}\\
O$^{13}$CS&4.1$\times$10$^{12}$&66.6$\pm$6.4&4.1$\times$10$^{-10}$&Warm component  &\citet{qui24}\\
OC$^{34}$S&9.5$\times$10$^{12}$&80.1$\pm$19.5&9.5$\times$10$^{-10}$&Warm component  &\citet{qui24}\\
\ce{SO2}&9.8$\times$10$^{12}$&62.5$\pm$29.1&9.8$\times$10$^{-10}$&Warm component  &\citet{qui24}\\
$^{34}$\ce{SO2}&1.1$\times$10$^{12}$&9.4$\pm$0.6&1.1$\times$10$^{-10}$&Cold component  &\citet{qui24}\\
\ce{H2S}&40$\times$10$^{12}$&47.2$\pm$5.2&4.0$\times$10$^{-9}$&Warm component  &\citet{qui24}\\
HDS&11.2$\times$10$^{12}$&10.8$\pm$2.2&1.12$\times$10$^{-9}$&Cold component  &\citet{qui24}\\
H$_{2}$$^{33}$S&2.5$\times$10$^{12}$&47$\pm$5&2.5$\times$10$^{-10}$&Warm component  &\citet{qui24}\\ 
\ce{D2S}&0.41$\times$10$^{12}$&11$\pm$2&4.1$\times$10$^{-11}$&Cold component  &\citet{qui24}\\ 
H$_{2}$$^{34}$S&11$\times$10$^{12}$&50.1$\pm$12&1.1$\times$10$^{-9}$&Warm component  &\citet{qui24}\\
CS&1.37$\times$10$^{15}$&110$\pm$12&9.8$\times$10$^{-8}$&Hot component  &\citet{bl95}\\
$^{13}$CS&13$\times$10$^{12}$&11.5$\pm$3&1.3$\times$10$^{-9}$&Cold component  &\citet{qui24}\\
C$^{33}$S&0.83$\times$10$^{12}$&12.6$\pm$2.7&8.3$\times$10$^{-11}$&Cold component  &\citet{qui24}\\
C$^{34}$S&3.3$\times$10$^{12}$&9.8$\pm$1.3&3.3$\times$10$^{-10}$&Cold component  &\citet{qui24}\\
HCS$^{+}$&0.44$\times$10$^{12}$&12.8$\pm$1.2&4.4$\times$10$^{-11}$&Cold component  &\citet{qui24}\\
NS&8.8$\times$10$^{12}$&8.6$\pm$0.3&8.8$\times$10$^{-10}$&Cold component  &\citet{qui24}\\
NS$^{+}$&0.10$\times$10$^{12}$&6.6$\pm$0.3&1.1$\times$10$^{-11}$&Cold component  &\citet{qui24}\\
\ce{H2CS}&9.3$\times$10$^{12}$&49.7$\pm$14.1&9.3$\times$10$^{-10}$&Warm component  &\citet{qui24}\\
HDCS&3.5$\times$10$^{12}$&15.7$\pm$0.7&3.5$\times$10$^{-10}$&Cold component  &\citet{qui24}\\
\ce{D2CS}&2.3$\times$10$^{12}$&12.8$\pm$1.5&2.3$\times$10$^{-10}$&Cold component  &\citet{qui24}\\
H$_{2}$C$^{34}$S&7.4$\times$10$^{12}$&10.2$\pm$1.2&7.4$\times$10$^{-10}$&Cold component  &\citet{qui24}\\
\ce{CH3SH}&4.8$\times$10$^{12}$&20.3$\pm$6.3&4.8$\times$10$^{-10}$&Cold component  &\citet{qui24}\\
CCS&2.8$\times$10$^{12}$&15.5$\pm$1.5&2.8$\times$10$^{-10}$&Cold component  &\citet{qui24}\\
CCCS&1.0$\times$10$^{12}$&12.7$\pm$3.4&1.0$\times$10$^{-10}$&Cold component  &\citet{qui24}\\
HSCN&7.0$\times$10$^{12}$&20.1$\pm$3.5&7.0$\times$10$^{-10}$&Cold component  &\citet{qui24}\\
t-HC(S)SH&(2.63$\pm$0.32)$\times$10$^{15}$&255$\pm$32&(2.53$\pm$0.68)$\times$10$^{-9}$&Hot component  & This study\\
\hline
\end{tabular}	
\end{adjustbox}
\label{tab:columndensity}
\end{table*}

\subsection{4.2 Comparison of the abundance of t-HC(S)SH with other S-bearing molecules}
Previous studies have shown that hot corino source A2 is a shelter of different S-bearing molecules, including different isotopes and deuterium \citep{bl95, qui24}. To understand the distributions of different S-bearing molecules, we compared our derived abundance of t-HC(S)SH with all the identified S-bearing molecules towards A2, which are listed in Table~\ref{tab:columndensity}. We observed that the abundance of S-bearing molecules towards A2 varied between $\sim$10$^{-11}$ and $\sim$10$^{-8}$. During chemical modelling, we observed that CS acts as a possible precursor of HC(S)SH, and the observed abundance of CS is approximately one order of magnitude higher than the abundance of t-HC(S)SH towards A2. Since the abundance of CS is higher than that of t-HC(S)SH, there is a high probability that t-HC(S)SH may form on the grain surface of A2 via reactions 1 and 2. From the Table~\ref{tab:columndensity}, we notice that the abundance of t-HC(S)SH is very close to that of SO, OCS, \ce{H2S}, HDS, H$_{2}$$^{34}$S, and $^{13}$CS. We also observe that the excitation temperatures of most of the S-bearing molecules are in the range of 7 K and 80 K, which indicates that these S-bearing species are emitted from the cold and warm regions of A2. We also observe that the excitation temperatures of both CS and t-HC(S)SH are above 100 K, which indicates that both molecules are emitted from the hot inner regions of A2. This indicates that both CS and t-HC(S)SH have chemical links. During chemical modelling, we also used gas phase reactions between HS and \ce{H2CS} to produce HC(S)SH (reaction 3) but we observed that those reactions did not produce HC(S)SH. After comparison, we observe that the abundance of \ce{H2CS} is approximately one order of magnitude lower than the abundance of t-HC(S)SH towards A2. This indicates that reaction 3 does not have the capability to produce HC(S)SH towards A2 in the gas phase. Previously, \citet{kou17} computed a chemical model of different S-bearing species and showed that both the gas phase and grain surface mechanisms are responsible for the production of different S-bearing molecules towards A2. Since the majority of S-bearing molecules are already detected towards A2, this source is an ideal target in the ISM to study more complex S-bearing molecules. 

\section{5. Conclusions}
In this article, we present the first detection of rotational emission lines of t-HC(S)SH towards A2. This is the first detection of this molecule in ISM. These results represent a major breakthrough in understanding sulphur chemistry in the ISM. Based on LTE spectral analysis, we found that the column density and excitation temperature of t-HC(S)SH towards A2 were (2.63$\pm$0.32)$\times$10$^{15}$ cm$^{-2}$ and 255$\pm$32 K, respectively. The fractional abundance of t-HC(S)SH with respect to \ce{H2} was (2.53$\pm$0.68)$\times$10$^{-9}$. We also computed a two-phase warm-up chemical model of HC(S)SH and concluded that t-HC(S)SH is formed in A2 via barrierless radical-radical reactions between CSSH and H on grain surfaces. The detection of t-HC(S)SH indicates a high chance of detecting the emission lines of other complex S-bearing molecules using ALMA in the near future. The detection and chemical modelling of t-HC(S)SH opens a new window for understanding sulphur chemistry in the ISM. We recommend the astrochemistry community to search for more high-intensity transitions of t-HC(S)SH towards A2 using the ALMA band 7 above a frequency of 300 GHz with a spectral span of $>$2 GHz.

\section*{Author information}
\text{Corresponding Author}: Arijit Manna\\
\text{E-mail: amanna.astro@gmail.com}\\
\text{ORCID}:\\
\text{Arijit Manna: 0000-0001-9133-3465}\\
\text{Sabyasachi Pal: 0000-0003-2325-8509}\\
\text{Notes: The authors declare no competing financial interest.}

\begin{acknowledgement}
We thank the anonymous reviewers for their helpful comments, which improved the manuscript. A.M. acknowledges the Swami Vivekananda Merit-cum-Means Scholarship (SVMCM) for financial support for this research. The chemically rich molecular emission spectra of NGC 1333 IRAS 4A2 are available on our \href{https://github.com/astrochemistry/NGC-1333-IRAS-4A}{GitHub} repository.
This paper makes use of the following ALMA data: ADS/JAO.ALMA\#2015.1.00147.S. ALMA is a partnership of ESO (representing its member states), NSF (USA), and NINS (Japan), together with NRC (Canada), MOST and ASIAA (Taiwan), and KASI (Republic of Korea), in cooperation with the Republic of Chile. The Joint ALMA Observatory is operated by ESO, AUI/NRAO, and NAOJ.

\end{acknowledgement}



\pagebreak 
\section{Graphical Abstract}
\begin{figure}
	\centering
	\includegraphics[width=1.0\textwidth]{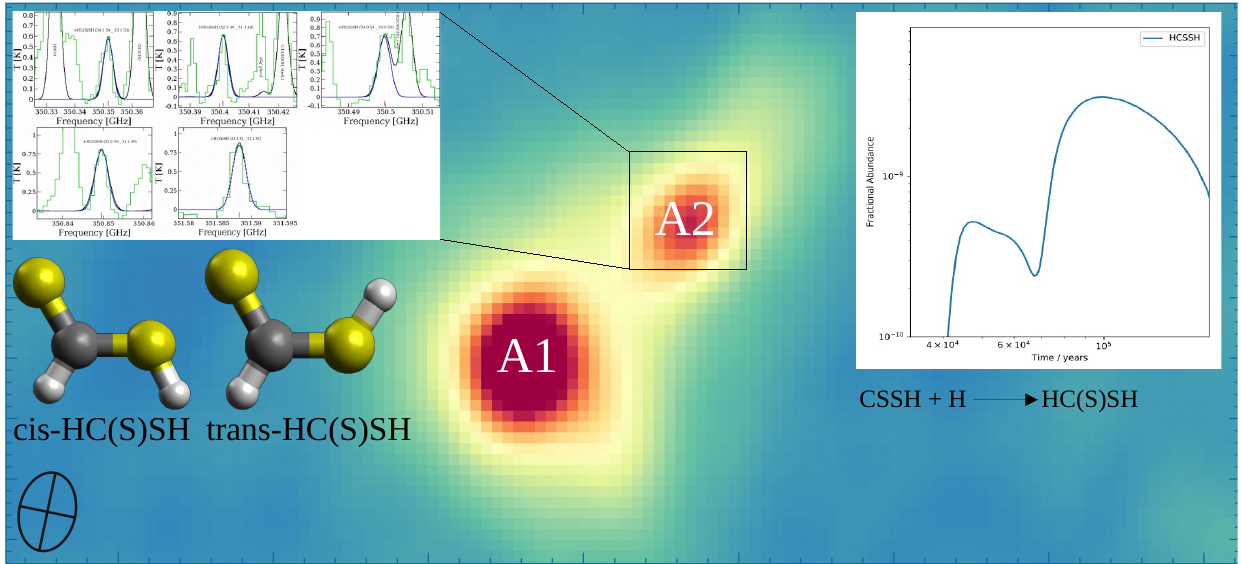}
\end{figure}
\end{document}